\documentclass[aps,pre,twocolumn,groupedaddress,superscriptaddress, showpacs]{revtex4-1}

\usepackage{graphicx}

\usepackage[usenames,dvipsnames]{color}

\usepackage{transparent}

\usepackage{tipa}

\usepackage{amsmath}

\usepackage{amsfonts}

\usepackage{amssymb}

\usepackage{dcolumn}

\usepackage[english]{babel}

\usepackage{array}

\begin{document}

\title{Usage leading to an abrupt collapse of connectivity}

\author{D. V. St\"ager}

  \email{staegerd@ethz.ch}

  \affiliation{Computational Physics for Engineering Materials, IfB, ETH Zurich, Wolfgang-Pauli-Strasse 27, CH-8093 Zurich, Switzerland}

\author{N. A. M. Ara\'ujo}

  \email{nmaraujo@fc.ul.pt}

  \affiliation{Departamento de F\'{\i}sica, Faculdade de Ci\^{e}ncias, Universidade de Lisboa, P-1749-016 Lisboa, Portugal, 
and Centro de F\'isica Te\'orica e Computacional, Universidade de Lisboa, P-1749-016 Lisboa, Portugal}

\author{H. J. Herrmann}

  \email{hans@ifb.baug.ethz.ch}

  \affiliation{Computational Physics for Engineering Materials, IfB, ETH Zurich, Wolfgang-Pauli-Strasse 27, CH-8093 Zurich, Switzerland}
  
  \affiliation{Departamento de F\'isica, Universidade Federal do Cear\'a, 60451-970 Fortaleza, Cear\'a, Brazil}

\begin{abstract}
Network infrastructures are essential for the distribution of resources such as electricity and water.
Typical strategies to assess their resilience focus on the impact of a sequence of random or targeted failures of network nodes or links.
Here we consider a more realistic scenario, where elements fail based on their usage.
We propose a dynamic model of transport based on the Bak-Tang-Wiesenfeld sandpile model where links fail after they have transported more than an amount $\mu$ (threshold) of the resource and we investigate it on the square lattice. As we deal with a new model, we provide insight on its fundamental behavior and dependence on parameters.
We observe that for low values of the threshold due to a positive feedback of link failure, an avalanche develops that leads to an abrupt collapse of the lattice. By contrast, for high thresholds the lattice breaks down in an uncorrelated fashion.
We determine the critical threshold $\mu^*$ separating these two regimes and show how it depends on the toppling threshold of the nodes and the mass increment added stepwise to the system. We find that the time of major disconnection is  well described with a linear dependence on $\mu$. Furthermore, we propose a lower bound for $\mu^*$ by measuring the strength of the dynamics leading to abrupt collapses.
\end{abstract}


\maketitle

\section{Introduction}
Economy increasingly relies on network-like infrastructures whose failures are costly such as power grids, water supply networks, public transportation systems, road networks, and the Internet \cite{CostPowerOutage1,CostWaterOutage1,CostHighwayCongestion}. Models to assess and quantify their resilience are more important than ever. The traditional approach monitors how sequences of failures of network nodes or links impact the global connectivity. For simplicity, these sequences have been drawn randomly \cite{ResilienceOfInternet,
RobustCoupledNetworks,ErrorAndAttackTolerance,NetworkRobustnessAndFragility}, based on topological properties (malicious attacks) \cite{MitigationMaliciousAttacks,BreakdownInternetAttack}, or according to the dynamics of cascading \cite{CatastrophicCascadesInterdependentNetworks,Cascadecontrol}. Different from that, we propose a model where links age, and thus they fail based on their cumulative usage. 

The dynamics of transport on a network can often be described by spatial load correlation. For example, a traffic jam in one avenue is likely to trigger congestion in neighboring roads. Similarly, one overloaded power station is typically surrounded by others working at full power \cite{Powershedding}. Also, when a reservoir overspills, it is very likely to trigger the spilling over of other reservoirs downstream \cite{Overspill}. To grasp such spatial and temporal correlation in a simple manner, we consider the Bak-Tang-Wiesenfeld (BTW) sandpile model \cite{BTW87}, where the iterative addition of sand grains on network nodes triggers avalanches of sand that propagate through the system.
The BTW model is among the simplest models exhibiting cascading dynamics which self-organize into a critical state \cite{ViewPointNuno,SandpileScaleFreeNetworks,ControlSelfOrgModelsSelfOrg}.
Because infrastructures are geographically embedded graphs, we will consider a square lattice, because the properties of the BTW and similar models on it have been studied extensively \cite{BTW87,BTW88,SOC,Grassberger90,Manna90,Pietronero94,Tebaldi99}. Power-law distributions of avalanches, as the ones predicted by the BTW model, have been observed in several physical networks such as electrical power grids \cite{BlackoutsSOC}, water reservoir networks \cite{Overspill}, and neural networks \cite{Beggs03122003,LucillaNeuronal}.

Only if no material is lost during avalanches, i.e., at dissipation rate $\varepsilon=0$, the dynamics is critical such that one finds a power-law avalanche size distribution \cite{Lauritsen96}. For $\varepsilon > 0$, there is an exponential cutoff in the avalanche size distribution at a characteristic size, such that the likelihood of avalanches larger than this characteristic size is negligible. The dynamics therefore is said to be subcritical.
Here, we only consider $\varepsilon > 0$ but, due to a positive feedback through link failures, we systematically find avalanches larger than this characteristic size. When links fail sufficiently fast, the lattice collapses abruptly at a certain time due to a single avalanche.
The size of this avalanche is of the size of the lattice due to a self-amplifying mechanism, and it is much larger than the size of any other observed avalanche. It is therefore an outlier similar to the so called ``Dragon-Kings'' found across various fields \cite{dragonkings}.

Our model, where links fail due to cumulative usage, has similarities with fracture, random fuse (RFM), \cite{randomfuse,Kahng88} and fiber bundle models (FBM) \cite{fiberbundlefirstpaper,fiberbundleoverview} where nodes (fuses and fibers) fail due to cumulative damage \cite{Durham97,Park05,LennartzSassinek13,Kumar04}. 
Due to spatial load correlation (RFM) and load redistribution (FBM), such systems undergo abrupt failure if the strength of nodes is narrowly distributed compared to an uncorrelated gradual fracture otherwise. In our model the analogy to the load is the usage of links which triggers link failure. Since avalanches are spatially embedded, there is spatial usage correlation. Furthermore, similar to the local load sharing FBM, the failure of links can immediately enhance the usage in their neighborhood, inducing a positive feedback of link failure. Depending on the strength of this positive feedback, we also find either an abrupt collapse or a gradual destruction.

This paper is organized as follows. In Sec. \ref{sec:Model} we introduce the model. Results are presented in Sec. \ref{sec:Results} and we draw conclusions in Sec. \ref{sec:Conclusions}.

\section{Model}\label{sec:Model}
We consider a square lattice with periodic boundary conditions i.e., we have $N=L^2$ nodes with initially ${2 L^2}$ links. 
Each node $i$ carries an amount of mass $z_i$ which can be transferred to other nodes through links.
Nodes topple if their mass $z_i$ is larger or equal the toppling threshold $z_c$, which we set equal to the initial degree of the nodes ($z_c=4$). A toppling of node $i$ leads to mass being distributed equally among its connected neighbors $j$. Every mass transfer from node $i$ to node $j$ is added to the usage $u_{ij}$ of the link in between. We have

$\begin{aligned}[t]
\textit{if} \ z_i \geq z_c:
\end{aligned}$
\begin{enumerate}
\item $\begin{aligned}[t] &\textit{for all nodes j linked to node i:} \\
&z_j \to z_j + \frac{z_i}{k_i}(1-\varepsilon), \\
&u_{ij} \to u_{ij} + \frac{z_i}{k_i}(1-\varepsilon),
\end{aligned}$
\item $\begin{aligned}[t]
  z_i \to 0,
\end{aligned}$
\end{enumerate}
where $k_i$ is the degree of the toppling node and the rate of dissipation $\varepsilon=0.01$, unless otherwise stated.
One toppling node can trigger the toppling of neighboring nodes what might lead to a cascade of topplings.
Each sequence of toppling nodes is considered an avalanche whose size $s$ is defined as the number of nodes that toppled at least once during the avalanche.
Initially every node is assigned a random mass $z_i$ uniformly distributed on $[0,z_c)$. Then increments of mass $\Delta z = 1$ are iteratively added to randomly chosen nodes, which eventually trigger the toppling of nodes.
Relaxation occurs on a much shorter time scale than external perturbations, i.e., only when an avalanche ends, the next increment of mass is added.
Before we allow links to fail, we first reach the stationary state where one finds a power-law avalanche size distribution truncated by an exponential cutoff at a characteristic avalanche size, as well known for this sandpile model with $\varepsilon>0$ \cite{Lauritsen96}. Note that this characteristic size is smaller than $L^2$ for the values of $L$ and $\varepsilon$ considered here.

After reaching the stationary state we set the time $t=0$, reset all $u_{ij}=0$, and introduce a failing threshold $\mu$ such that after each toppling all links with $u_{ij}>\mu$ fail and are removed, what decreases the degrees $k_i$ and $k_j$. In one time unit each node receives on average one unity of mass. The final state is reached when all links have failed.

\section{Results}\label{sec:Results}
Depending on the failing threshold $\mu$ one finds two different regimes separated by a critical threshold $\mu^*$ in the thermodynamic limit and separated by a transition zone $\mu \approx \mu^*_{eff}$ for finite lattices. For simplicity, we will refer to the two regimes only as $\mu < \mu^*$ and $\mu > \mu^*$.

\begin{figure}[]
\begin{center}
	\includegraphics[width=\columnwidth]{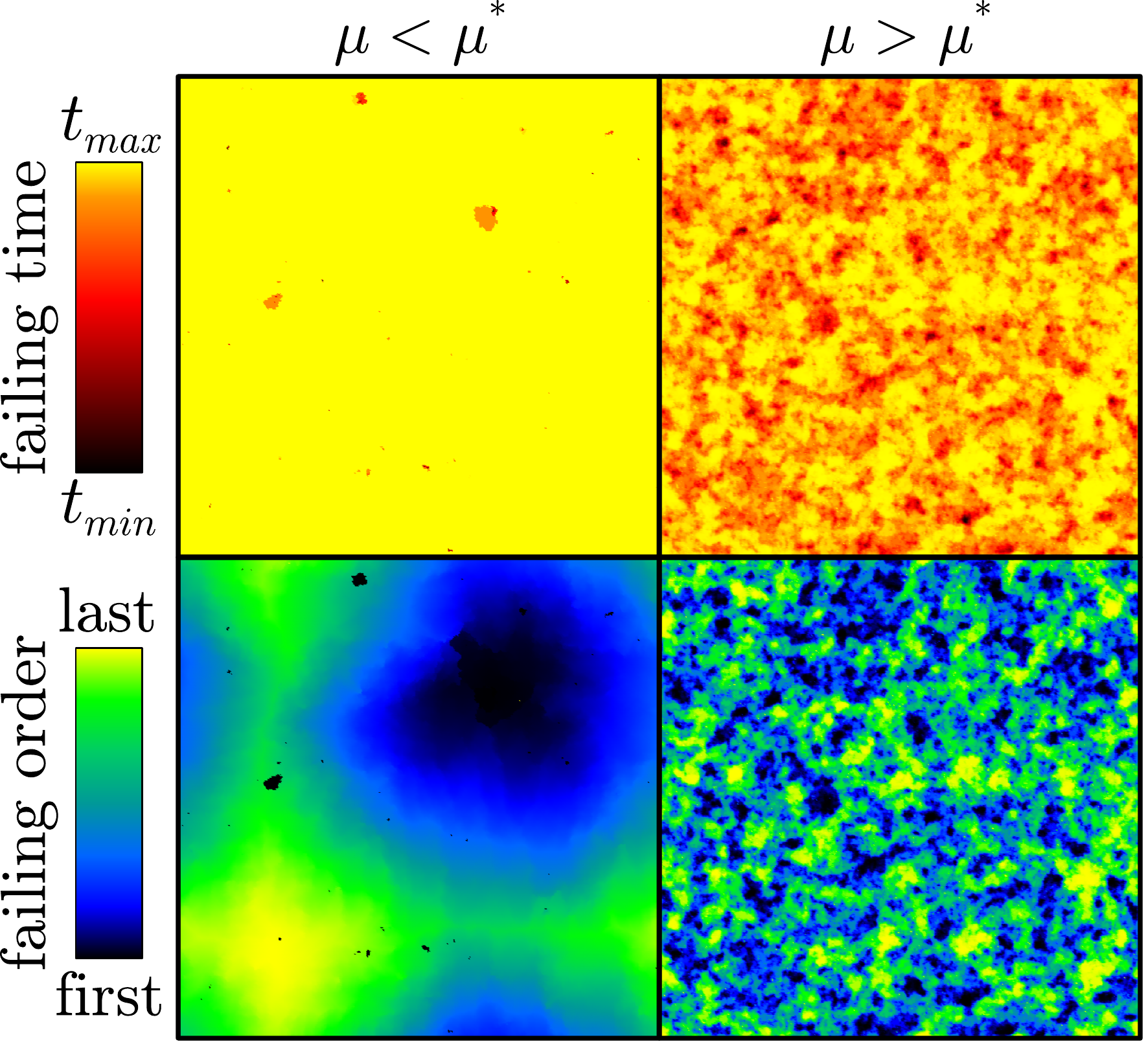}
\end{center}
\caption{
\label{fig:explanatory}
(Color online) \textbf{Link failing time (top) and failing order (bottom) for the two different regimes for a single sample.} For $\mu < \mu^*$ $(\mu = 100)$ after the failure of a minor fraction of links, one devastating avalanche destroys the entire network at once (top left) with the damage propagating radially outwards (bottom left). By contrast, for $\mu > \mu^*$ $(\mu = 10^5)$ each avalanche only destroys a minor fraction of links and no long range correlation in failing time (top right) and order (bottom right) can be observed. $t_{min}$ and $t_{max}$ denote the time of the first and the last failed link respectively. Results were obtained on a square lattice of $L = 512$, with dissipation rate $\varepsilon = 0.01$.
}
\end{figure}

Let us define the number of links failed during an avalanche as the avalanche damage $d$, and its maximum within the same sample as $d_{max}$.
For $\mu<\mu^*$, one single macroscopic avalanche destroys almost every link of the lattice, i.e., one finds (details below)
\begin{equation}\label{eq:limitto1}
\textit{for}\ \mu<\mu^*: \ \lim_{L\to \infty} \frac{d_{max}}{2L^2} = 1.
\end{equation}
We denote such an avalanche as a \emph{devastating avalanche}.
Figure \ref{fig:explanatory} (top left) exemplarily shows the time when each link fails for one configuration. After the failing of minor parts due to avalanches with small $d$, the whole remaining lattice is destroyed at once by one single devastating avalanche with $d_{max}$. 
Initially links start to fail in one area and failing then spreads outwards. This can be seen in Fig.\ \ref{fig:explanatory} (bottom left) which shows the order of failing of the links for the same configuration as in Fig.\ \ref{fig:explanatory} (top left). Due to the radial propagation of the failing, one finds a strong long range spatial correlation in the failing order of links.
By contrast, for $\mu > \mu^*$, avalanches only cause the failure of a small number of links, such that
\begin{equation}
\textit{for}\ \mu>\mu^*: \ \lim_{L\to \infty} \frac{d_{max}}{2L^2} = 0.
\end{equation}
One observes a short range correlation in failing time and failing order as seen in Fig.\ \ref{fig:explanatory} (right).

In the following subsections we will describe both regimes $\mu<\mu^*$ and $\mu>\mu^*$ in detail, determine $\mu^*$, investigate the time of major disconnection, show the role of the toppling threshold of nodes and the mass increment added stepwise to the system, and discuss how to prevent an abrupt collapse. To quantify the process of disconnection of the nodes, we define the quantity $S$ as the fraction of nodes belonging to the largest connected component, such that a fully connected lattice has $S=1$ and a fully disconnected one $S=0$.

\subsection{Abrupt collapse: $\mu < \mu^*$}\vspace*{-1.5mm}

For $\mu < \mu^*$, links only participate in a small number of avalanches till they fail. The difference in $u_{ij}$ of neighboring links remains low till they fail and since the failing threshold is equal for all links, it is likely that neighboring links fail due to the same toppling or consecutive ones. 
Given that, there are two mechanisms which are responsible for a devastating avalanche. First, as links start to fail, the effective degree of the nodes decreases. The smaller the effective degree of a toppling node, the more mass is transported through each of its links. So links are more likely to fail in the neighborhood of a previously failed link. Second, when an avalanche starts to destroy many links consecutively, the failing spreads out. Mass is then accumulated next to the border of the destroyed area triggering many more toppling events and therefore even further failing of links.
These two mechanisms help sustaining the avalanche which therefore leads to an abrupt collapse of the lattice.
\begin{figure}[]
\begin{center}
	\includegraphics[width=\columnwidth]{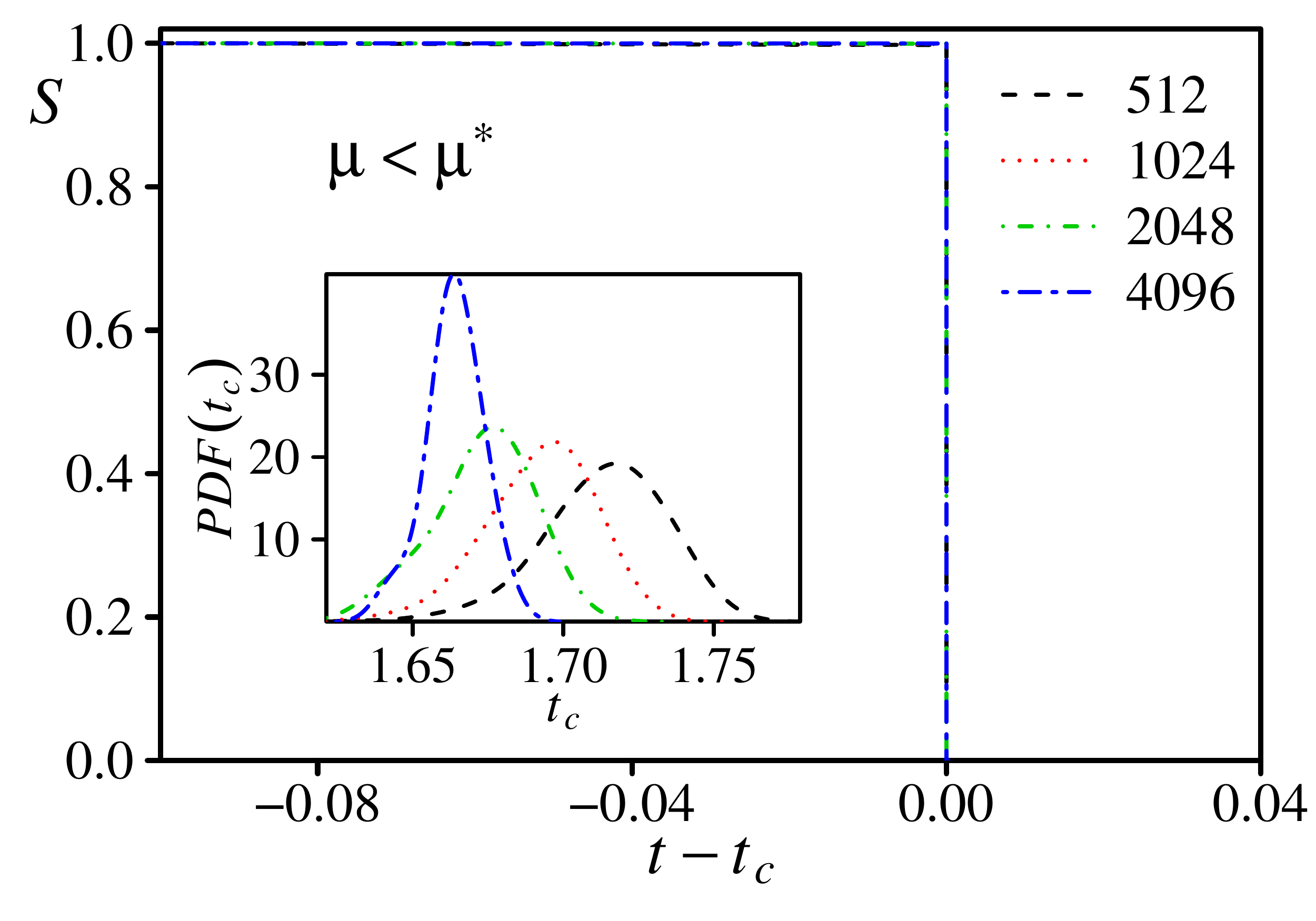}
\end{center}
\caption{
\label{fig:discontinuous}(Color online) \textbf{Abrupt collapse for low failing thresholds.} For $\mu < \mu^*$ $(\mu = 100)$ due to a single devastating avalanche, the fraction of nodes belonging to the largest connected component decreases from $S \approx 1$ to $S \approx 0$ discontinuously. The time $t_c$ when this avalanche occurs is approximately Gaussian distributed (inset) with a mean and standard deviation which decrease with $L$. For $L=\{512,1024,2048,4096\}$ the number of samples is $\{1000,250,125,25\}$. $\varepsilon=0.01$.
}
\end{figure}

Figure \ref{fig:discontinuous} shows the time evolution of the fraction $S$ of nodes belonging to the largest connected component, where a discontinuous transition is observed. We denote $t_c$ as the time of the avalanche which leads to the largest decrease in $S$. Thus, in the regime $\mu<\mu^*$, $t_c$ is the time of the devastating avalanche. This time follows approximately a Gaussian distribution with a mean and standard deviation that decrease with $L$ (inset of Fig.\ \ref{fig:discontinuous}).

\begin{figure}[]
\begin{center}
	\includegraphics[width=\columnwidth]{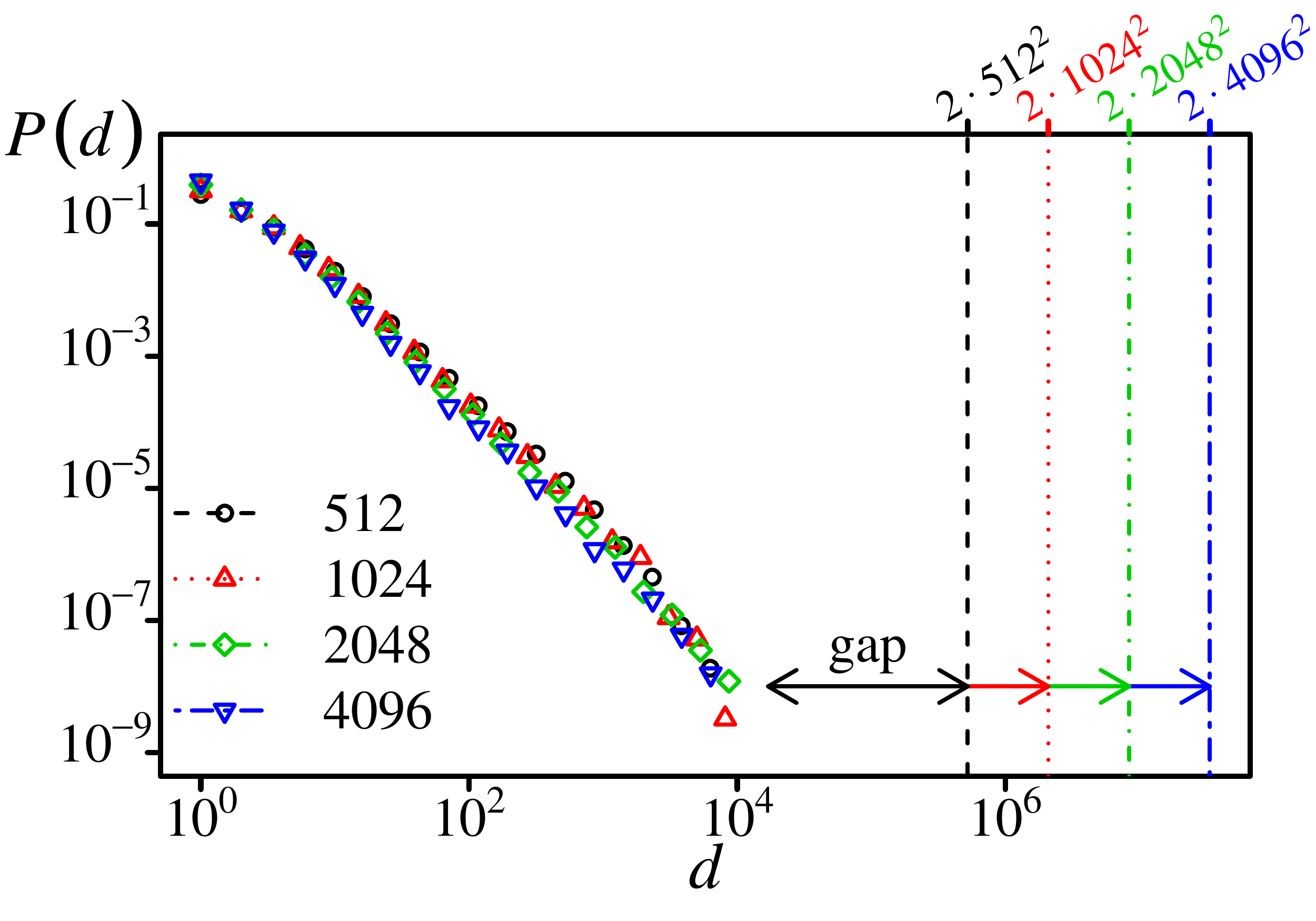}
\end{center}
\caption{
\label{fig:killsizeslowmu}
(Color online) \textbf{Avalanche damage distribution.} Damage probability distributions of prior avalanches (symbols) with $d \geq 1$ and damage due to the devastating avalanche (dashed lines) for $\mu < \mu^* (\mu = 100)$. Note the gap between the damage of prior avalanches and the devastating avalanche. For $L=\{512,1024,2048,4096\}$ the number of samples is $\{1000,250,125,25\}$. $\varepsilon=0.01$.
}
\end{figure}

We denote avalanches which occur before the devastating avalanche as \emph{prior avalanches}. The avalanche damage probability distribution of prior avalanches does not change significantly with system size and the largest damage due to a prior avalanche is small compared to the damage of the devastating avalanche as shown in Fig.\ \ref{fig:killsizeslowmu}. We find for $\mu=100$ that the fraction of links destroyed by prior avalanches scales with $L^{-\kappa}$, where $\kappa = 1.0 \pm 0.1$, and goes to zero in the thermodynamic limit as for all $\mu<\mu^*$. This justifies the limit given by Eq.\ (\ref{eq:limitto1}).

To grasp the dynamics as we approach the devastating avalanche, we investigated the size distribution of prior avalanches for $\mu = 100$. Long before $t_c$, when no link failed yet, we 
observe that the probability $p(s)$ that an avalanche of size $s$ occurs follows a power-law behavior $p(s)\sim s^{-\tau}$ with $\tau \approx 1$ as for the BTW model \citep{BTW87} and an exponential cutoff at the upper end, as expected in the presence of dissipation. When $t$ gets closer to $t_c$, the probability $p(s)$ for $s$ in the range of the cutoff and even above, where it was practically zero before, increases. By dividing the avalanches into destructive ones ($d>0$) and non-destructive ones ($d=0$), one finds that the increased probabilities $p(s)$ for large $s$ only comes from destructive avalanches. This confirms that link failing can amplify avalanches.

Note that generally link failing can either amplify or inhibit the strength of avalanches. Even though for the set of parameters considered here mostly the amplifying effect dominates, one should nevertheless be aware that link failing can also stop an avalanche, mainly when after the failing of links a node,  that is about to topple, is isolated. But the influence of this inhibiting effect is only noticeable for very low $\mu$ and decreases with system size.

\subsection{Gradual destruction: $\mu > \mu^*$}

\begin{figure}[]
\begin{center}
	\includegraphics[width=\columnwidth]{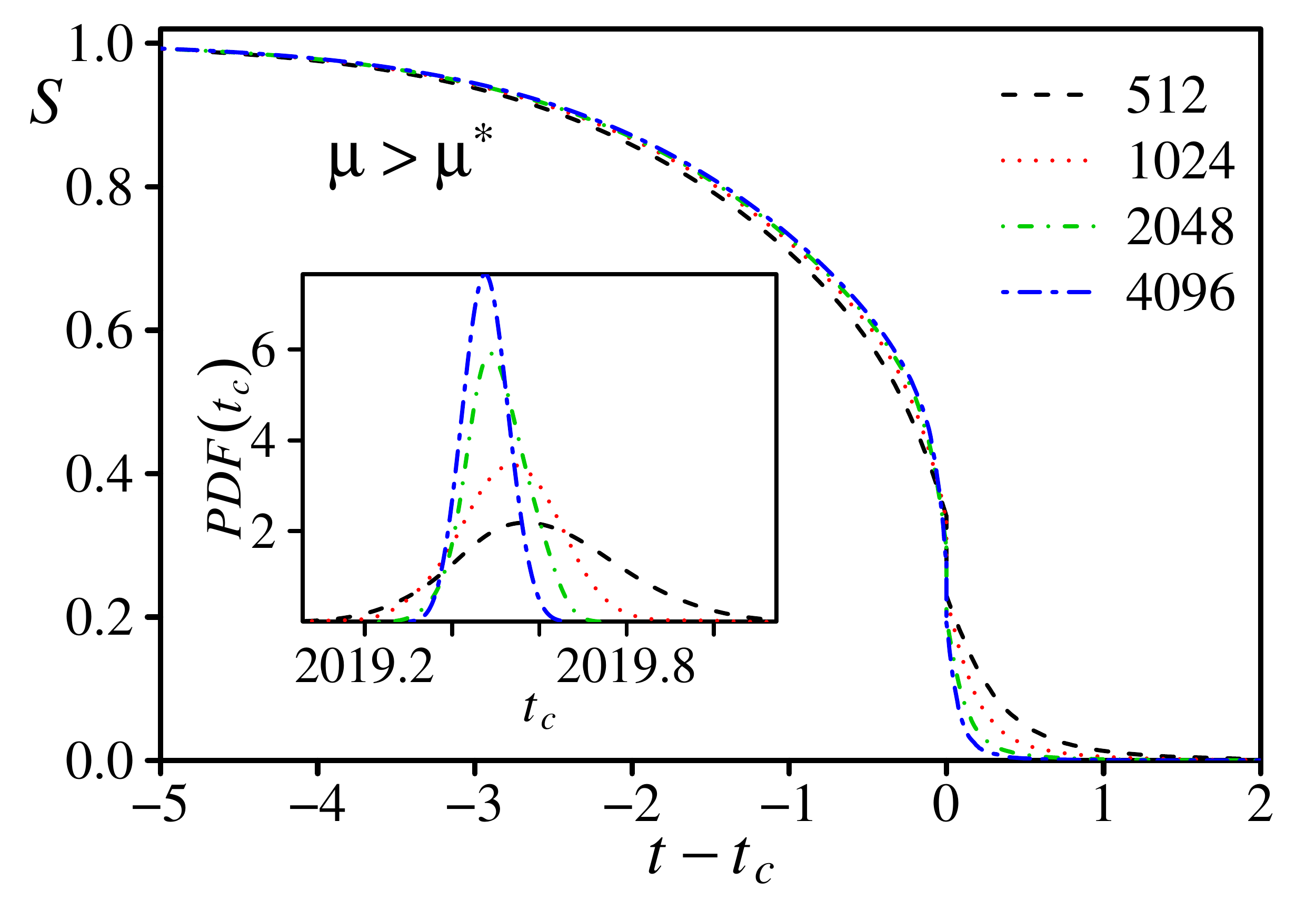}
\end{center}
\caption{
\label{fig:continuous}(Color online) \textbf{Gradual destruction for high failing thresholds.} For $\mu > \mu^*$ $(\mu = 10^5)$ the fraction of nodes belonging to the largest cluster $S$ decreases continuously in time. $t_c$ is the time at which the biggest jump $\Delta S$ due to a single avalanche occurs and follows approximately a Gaussian (inset) with a mean and standard deviation which decrease with $L$. For $L=\{512,1024,2048,4096\}$ the number of samples is $\{1000,250,125,25\}$. $\varepsilon=0.01$.
}
\end{figure}

\begin{figure}[]
\begin{center}
	\includegraphics[width=\columnwidth]{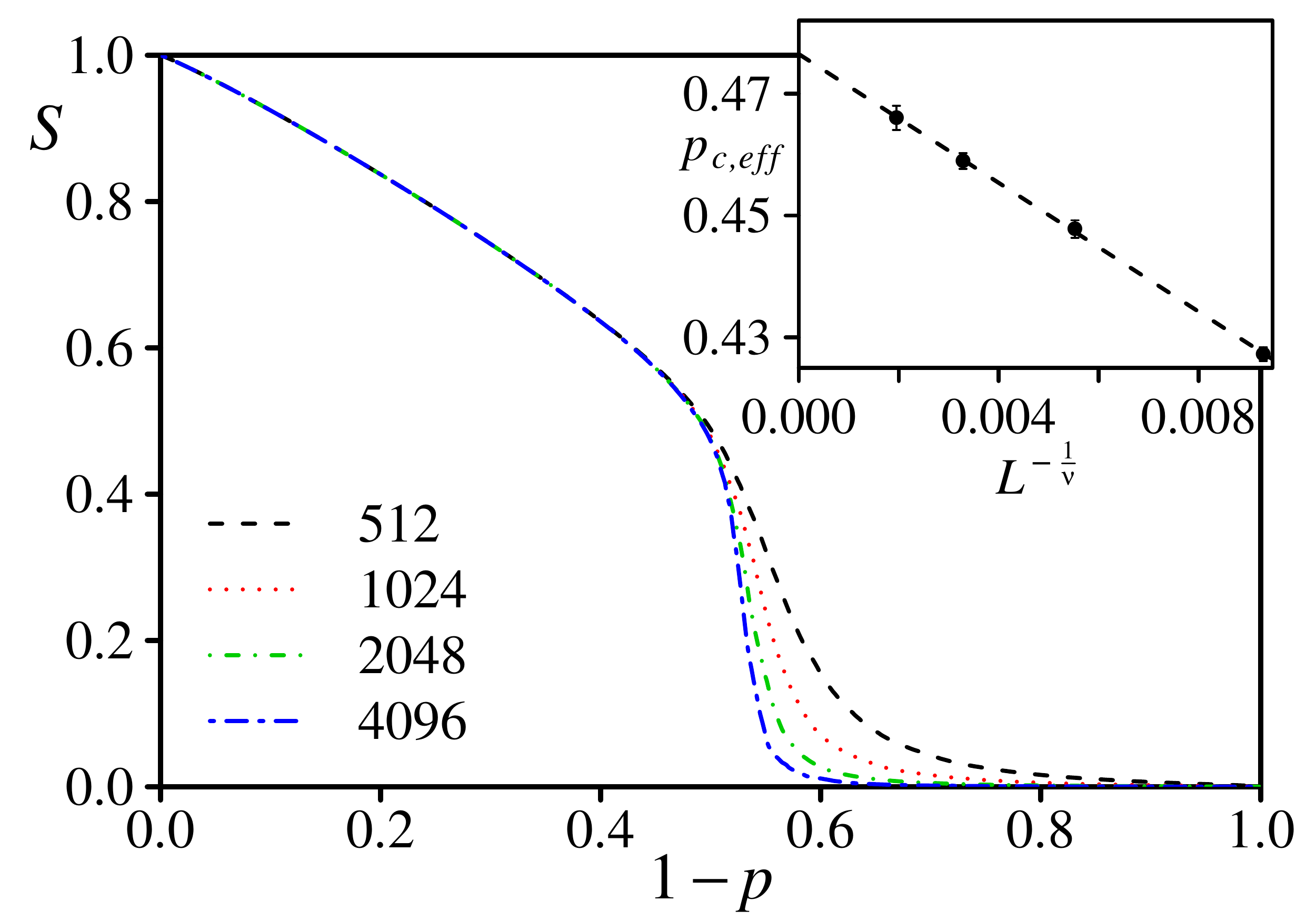}
\end{center}
\caption{
\label{fig:Svsqlargemu}
(Color online) \textbf{S versus fraction of failed links.} Fraction of nodes belonging to the largest cluster $S$ versus fraction of failed links $1-p$ for $\mu > \mu^* (\mu = 10^5)$ with dissipation $\varepsilon=0.01$. Inset: Scaling of the effective critical thresholds $p_{c,eff}$ with $\nu=4/3$ leads to $p_c=0.476 \pm 0.002$ in the thermodynamic limit. For $L=\{512,1024,2048,4096\}$ the number of samples is $\{1000,250,125,25\}$. $\varepsilon=0.01$.
}
\end{figure}

\begin{figure}[]
\begin{center}
	\includegraphics[width=\columnwidth]{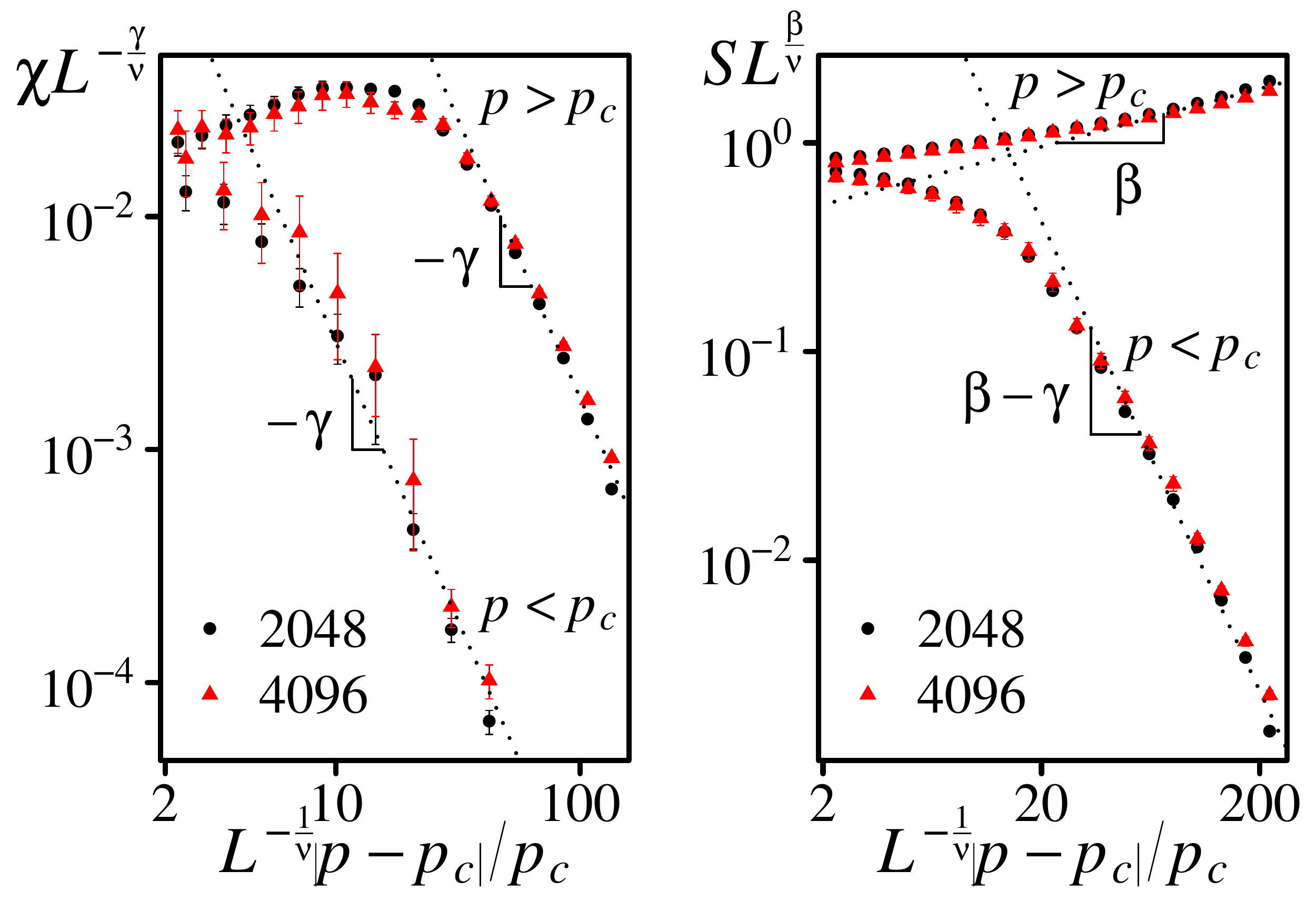}
\end{center}
\caption{
\label{fig:universalityclass}
(Color online) \textbf{Universality class of random percolation.} Finite-size scaling of the second moment of the cluster size distribution $\chi$ and fraction of total nodes belonging to the largest cluster $S$ for $\mu > \mu^* (\mu = 10^5)$. Our simulation data agrees with the scaling behavior known from random percolation $\chi L^{-\frac{\gamma}{\nu}} \sim L^{\frac{1}{\nu}}|p-p_c|/p_c$ with $\chi \sim |p-p_c|^{-\gamma}$ and $S L^{\frac{\beta}{\nu}} \sim L^{\frac{1}{\nu}}|p-p_c|/p_c$ with $S \sim (p-p_c)^{\beta}$ for $p>p_c$ respectively $S \sim (p_c-p)^{\beta-\gamma}$ for $p<p_c$. For $L=\{2048,4096\}$ the number of samples is $\{125,25\}$. $\varepsilon=0.01$.
}
\end{figure}

The higher the failing threshold, the more mass transport does it take to destroy a link and the larger will be the fluctuations in $u_{ij}$ of neighboring links when the links are close to failure. Therefore it is less probable for higher failing thresholds that neighboring links fail in the same avalanche.
For $\mu > \mu^*$, no avalanche dominates and $S$ decreases gradually with time as shown in Fig.\ \ref{fig:continuous}. The largest change in $S$ decreases with the system size and the transition is continuous for an infinite lattice. Again we find that $t_c$ is approximately Gaussian distributed (inset of Fig.\ \ref{fig:continuous}). In the thermodynamic limit we find for $\mu=10^5$ that $t_c=2019.46 \pm 0.02$ with the assumption $(t_c(L)-t_c) \sim L^{-\alpha}$ with $\alpha = 0.87$.
From the behavior of $S$ and the second moment of the cluster-size distribution $\chi$ versus the fraction of failed links $1-p$, where $p$ is the fraction of remaining links, we find that the transition is in the universality class of random percolation with critical exponents $\nu=4/3$, $\beta=5/36$, and $\gamma=43/18$. Figure \ref{fig:Svsqlargemu} shows $S$ versus $1-p$ and the effective critical threshold $(p_{c,eff}-p_c) \sim L^{-\frac{1}{\nu}}$ in the inset. Furthermore we show the finite-size scaling of $S$ and $\chi$ in Fig. \ref{fig:universalityclass}.

\subsection{At the transition: $\mu^*$}

\begin{figure}[]
\begin{center}
	\includegraphics[width=\columnwidth]{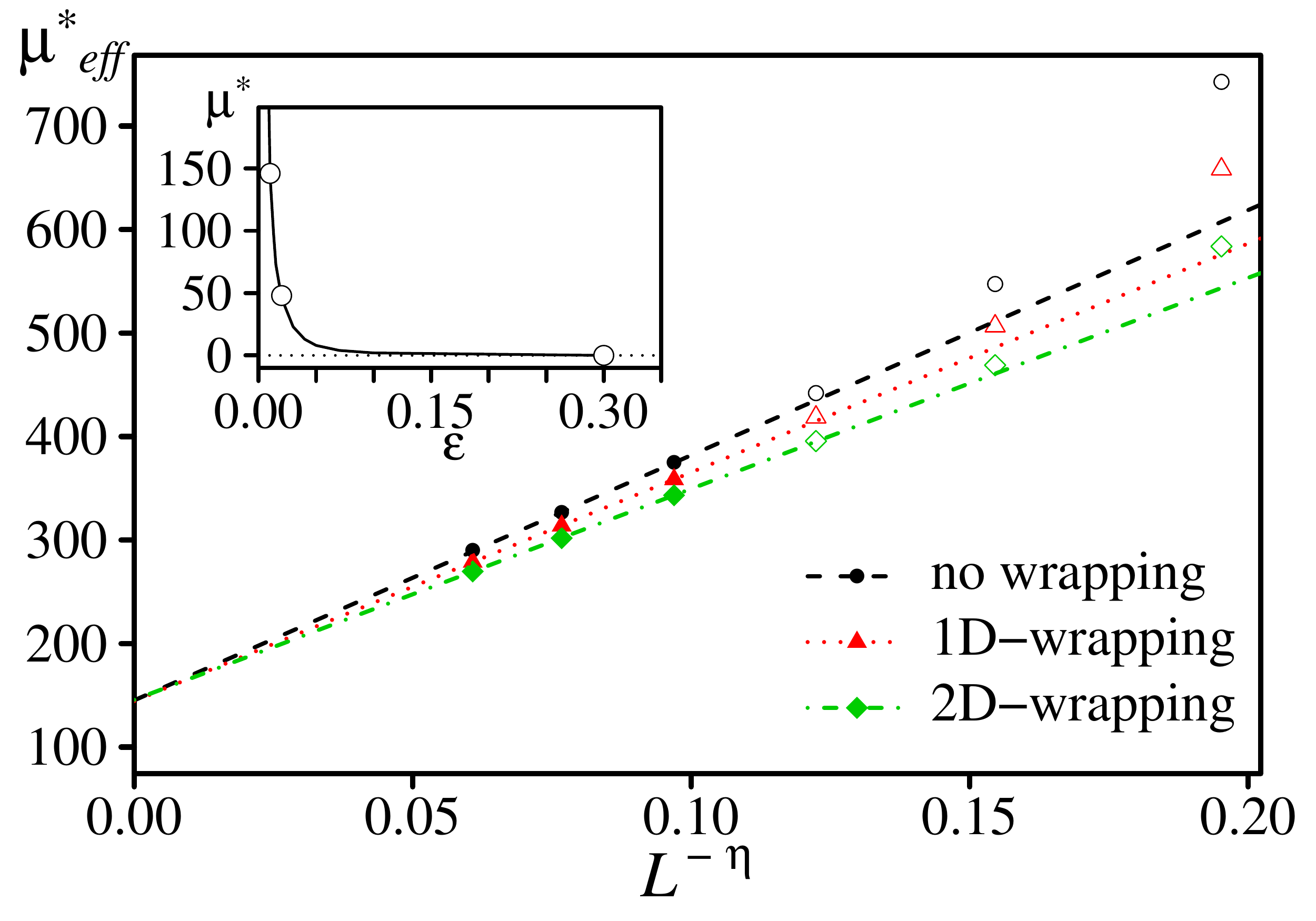}
\end{center}
\caption{
\label{fig:mustar}
(Color online) \textbf{Extrapolation to find the critical threshold.} Extrapolation of different estimators of $\mu^*_{eff}$ for $\varepsilon=0.01$ and $\eta = 0.34 \pm 0.01$.  Linear regressions (lines) for  $L = \{1024,2058,4096\}$ (solid symbols). Significant deviation is observed for small system sizes $L=\{128,256,512\}$ (open symbols). Error bars are below $0.6\%$. Inset: values of $\mu^*$ for $\varepsilon = \{0.01,0.02\}$ and a guide to the eyes (solid line) with the upper limit of dissipation $\varepsilon = 0.30 \pm 0.01$ where one only observes a devastating avalanche for $\mu=0$. For $L=\{128,256,512,1024,2048,4096\}$, measurements were performed in steps of $\Delta \mu = 10$ with number of samples being $\{3 \cdot 10^4,3 \cdot 10^4,7500,8000,1200,500\}$.
}
\end{figure}

In the regime of abrupt collapse the links destroyed by the devastating avalanche form a connected cluster that wraps around the lattice, which we denote as \emph{wrapping damage cluster}. One can not find this in the regime of gradual destruction. Thus, in the thermodynamic limit, the probability to find a wrapping damage cluster is a step function at $\mu=\mu^*$.
We use this fact to find several estimators for the effective threshold $\mu^*_{eff}$ to extrapolate to the critical threshold $\mu^*$.
As $\mu^*_{eff}$ we take the value of $\mu$ where the probability to find a wrapping damage cluster is equal to $1/2$.
In particular, we use three different estimators based on the probability of no wrapping, wrapping along only one direction (1D-wrapping), and wrapping in both directions (2D-wrapping).
We assume that all three estimators of $\mu^*_{eff}$ scale with lattice size as $L^{-\eta}$, with the same $\eta > 0$ and find the best value for $\eta$ such that the linear fitting $\mu^*_{eff}=\mu^*+cL^{-\eta}$ intercepts the origin at the same value $\mu^*$ for different estimators. For dissipation $\varepsilon=0.01$ we find $\eta = 0.34 \pm 0.01$ which results in $\mu^*=146 \pm 6$ as seen in Fig.\ \ref{fig:mustar}. For dissipation $\varepsilon=0.02$ we find $\eta = 0.29 \pm 0.01$ with $\mu^*=48 \pm 1$. With increasing dissipation, the transition between a discontinuous abrupt collapse and a gradual destruction is shifted to lower $\mu$ and for $\varepsilon > 0.30 \pm 0.01$ no devastating avalanche is observed even for $\mu = 0$ as seen in the inset of Fig.\ \ref{fig:mustar}.

For finite systems the used estimators for the effective critical thresholds $\mu^*_{eff}$ are upper bounds to $\mu^*$ as seen in Fig.\ \ref{fig:mustar}.
By checking for a wrapping damage cluster one can not give a lower bound for $\mu^*$, but we found a different approach to do so.
We analyze how powerful devastating avalanches are for a certain lattice size and out of this predict if such an avalanche would evolve to be powerful enough to destroy an infinite system. If we find that also an infinite system would collapse abruptly, we have found a lower bound for $\mu^*$.
We quantify the power of an avalanche in the following way.
Let the first toppling of an avalanche be its first step $l=1$. At each step $l$, all nodes with mass $z_i \geq z_c$ topple. 
We define the power $\Psi_l$ as the total mass $\mathcal{M}_l$ on toppling nodes divided by the number of toppling nodes $\mathcal{N}_l$ at step $l$, i.e., 

\begin{equation}
\Psi_l = \frac{\mathcal{M}_l}{\mathcal{N}_l}.
\end{equation}

\begin{figure}[]
\begin{center}
	\includegraphics[width=\columnwidth]{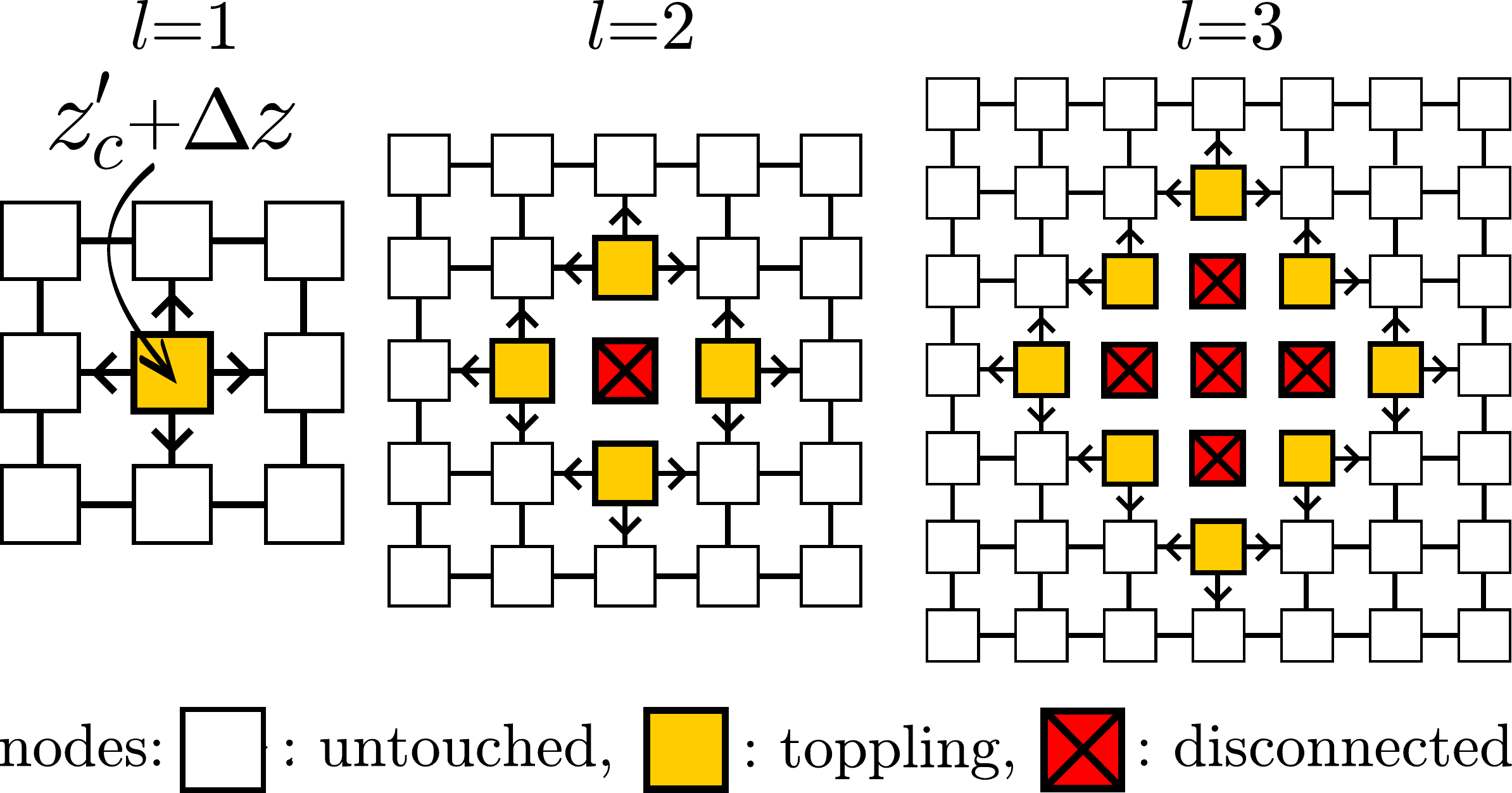}
\end{center}
\caption{
\label{fig:powersketch}
(Color online) \textbf{Devastating avalanche in the worst case scenario.} First three steps of the devastating avalanche when initially all nodes have mass $z_i = z_c' < z_c$, where $z_c'$ is arbitrarily close to $z_c$, and links fail after the first transport ($\mu=0$). Every node that receives mass topples and every link used is removed.
}
\end{figure}

For a devastating avalanche to occur in an infinite system, its power needs to grow to a level at which its further progress is guaranteed.
Such an avalanche will reach a constant level of power, at which mass accumulation due to the spreading of link failure balances out the suppression by dissipation. To clarify this, we discuss in detail the worst case scenario, i.e., where links fail after the first passage of mass ($\mu=0$) and where initially all nodes are filled completely, i.e., $z_i = z_c' < z_c$, where $z_c'=z_c-\delta$, and $\delta \to 0$. 
The first increment of mass $\Delta z=1$ will lead to a toppling since $z_c' + \Delta z\geq z_c$, such that at step $l=1$, as seen in Fig.\ \ref{fig:powersketch} (left), we have
\begin{eqnarray}
\begin{aligned}
\mathcal{M}_1 &= z_c' + \Delta z, \\ \mathcal{N}_1 &= 1, \\
\Psi_1 &= \frac{\mathcal{M}_1}{\mathcal{N}_1} = z_c' + \Delta z. \\
\end{aligned}
\end{eqnarray}
Fig.\ \ref{fig:powersketch} (middle) shows how after the toppling of this first node, all its links are removed (since $\mu=0$) and all its neighbors topple at step $l=2$. The total mass on toppling nodes at $l=2$ is equal to the mass $(1-\varepsilon)\mathcal{M}_1$ shed from the first toppled node plus the mass $\mathcal{N}_2 z_c'$ that beforehand was on the nodes, i.e. \
\begin{eqnarray}
\begin{aligned}
\mathcal{M}_2 &= (1-\varepsilon)\mathcal{M}_1+\mathcal{N}_2 z_c', \\ \mathcal{N}_2 &= 4.
\end{aligned}
\end{eqnarray}
We can now express $\Psi_2$ as
\begin{eqnarray}
\begin{aligned}
\Psi_2 &= \frac{\mathcal{M}_2}{\mathcal{N}_2} = \frac{(1-\varepsilon)\mathcal{M}_1+\mathcal{N}_2 z_c'}{\mathcal{N}_2} \\ &= (1-\varepsilon)\Psi_1\frac{\mathcal{N}_1}{\mathcal{N}_2}+ z_c'.\\
\end{aligned}
\end{eqnarray}
From this we find the iterative formula for $l>2$
\begin{eqnarray}
\begin{aligned}
\Psi_{l|l>2} &= (1-\varepsilon)\Psi_l\frac{\mathcal{N}_{l-1}}{\mathcal{N}_l}+ z_c',
\end{aligned}\label{eq:psiiterative}
\end{eqnarray}
where the number of toppling nodes increases by four each step (compare $l=2$ and $l=3$ in Fig.\ \ref{fig:powersketch}), i.e.,
\begin{eqnarray}
\begin{aligned}
\mathcal{N}_{l|l>1}=4(l-1).
\end{aligned}
\end{eqnarray}
We are only interested in the limit $l \to \infty$ and we use that for $l \gg 1$  
\begin{eqnarray}
\begin{aligned}
\frac{\mathcal{N}_{l-1}}{\mathcal{N}_l}=\frac{4(l-2)}{4(l-1)}\approx 1,
\end{aligned}
\end{eqnarray}
and simplify Eq.\ (\ref{eq:psiiterative}) to
\begin{eqnarray}
\begin{aligned}
\Psi_l &= (1-\varepsilon)\Psi_{l-1}  + z_c'.
\end{aligned}
\end{eqnarray}
One finds that $\Psi_{l}$ converges to its fixed point
\begin{eqnarray}
\begin{aligned}
\Psi_{l} &\xrightarrow{l \rightarrow \infty} \frac{z_c'}{\varepsilon} < \frac{z_c}{\varepsilon},
\end{aligned}
\end{eqnarray}
i.e., the most devastating avalanche reaches a constant power-level strictly smaller than $z_c/\varepsilon$.

\begin{figure}[]
\begin{center}
	\includegraphics[width=\columnwidth]{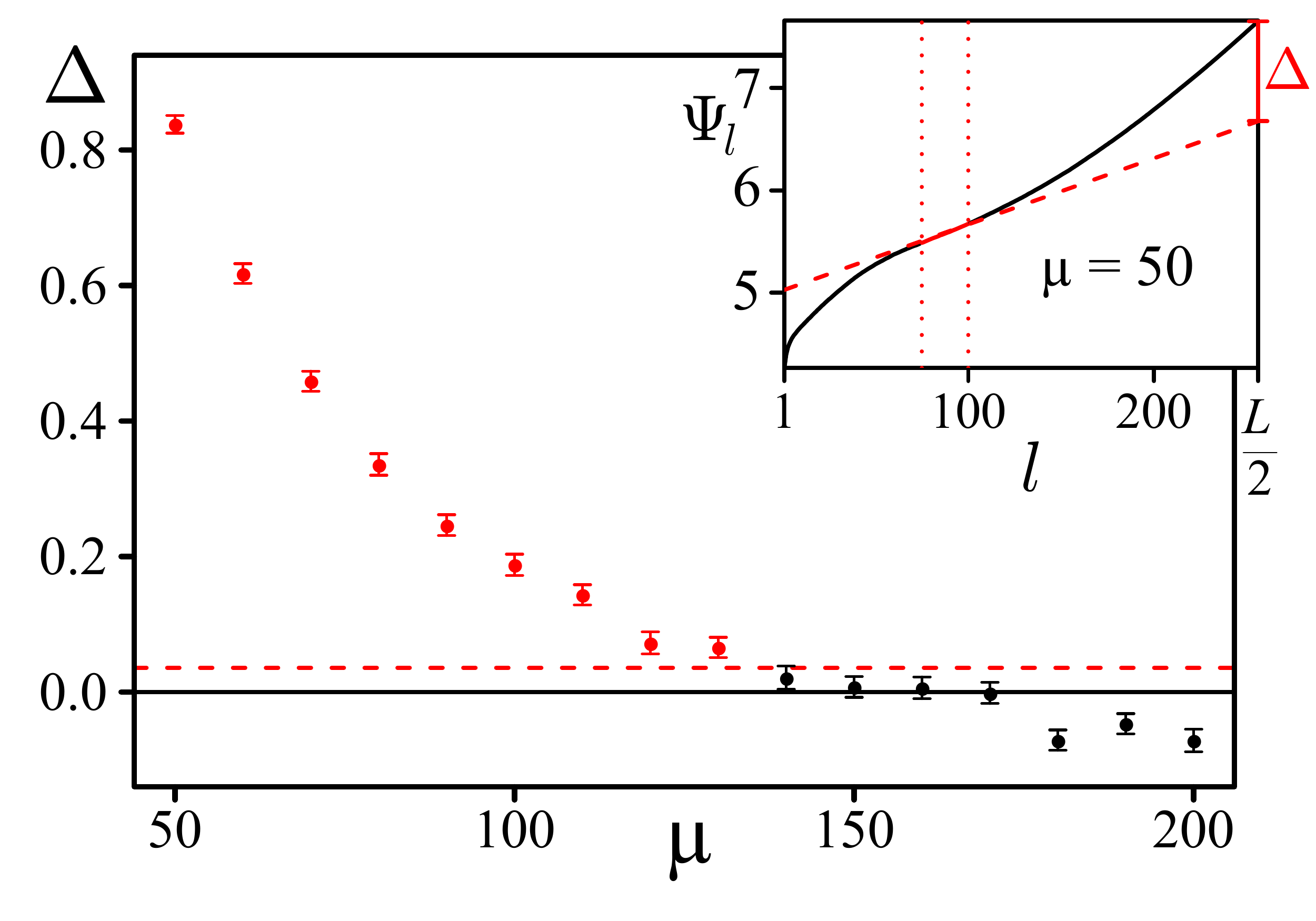}
\end{center}
\caption{
\label{fig:avpower}
(Color online) \textbf{Predicting a lower bound for the critical threshold.} Our measure $\Delta$ for super-linear growth versus $\mu$ for $L = 512$, averaged over $10^5$ samples. With a statistical test (Z-test) which assumes with the so called null hypothesis $H_0$ that $\Delta$ is Gaussian distributed with mean $\Delta_0=0$ and standard deviation equal to the average standard error of the data $\sigma_0=0.0154$, one expects $\Delta \geq 0.0358$ (see dashed line) only with probability $0.01$. For $\mu \leq 130$ we detect that there is significant super-linear growth in power ($\Delta=0.0658$ for $\mu = 130$ and probability $<10^{-5}$ that $\Delta \geq 0.0658$ under $H_0$). Inset: Power $\Psi_l$ of devastating avalanche versus step $l$ for $\mu=50$ averaged over $10^5$ samples. Tangent to $\Psi_l$ at the first inflection point $l = 87$ (dashed), fitted in the region $74 \le l \le 100$ (vertical dotted lines). Independently of $\mu$, we define $\Delta$ as the difference at $l=L/2$ between $\Psi_l$ and its tangent, where the tangent is incident to $\Psi_l$ at $l=87$. $\varepsilon=0.01$.
}
\end{figure}

For finite size lattices we are only interested in the power for $1\leq l \leq L/2$, where we are sure that the avalanche does not interact with itself wrapping around the lattice.
In simulations for $\mu < \mu^*$ we observe that the power of the devastating avalanche averaged over many configurations has a phase of super-linear increase as shown for $\mu=50$ in the inset of Fig.\ \ref{fig:avpower}. This shows that the rate of mass accumulation increases and the positive feedback of link failure gets stronger. The power of these avalanches will increase up to where dissipation balances out with mass accumulation, as discussed before. Every $\mu$ for which we can detect this super-linear power increase serves therefore as a lower bound for $\mu^*$.
We define
\begin{equation}
\Delta = \Psi_{L/2}-\tilde{\Psi}(L/2),
\end{equation}
where $\tilde{\Psi}(l)$ is the tangent to $\Psi_l$ that is incident at $l=87$ which is the first inflection point of $\Psi_l$ for $\mu=50$ (see dashed line in the inset of Fig.\ \ref{fig:avpower}).
We show for $L = 512$ in Fig.\ \ref{fig:avpower} that there is significant super-linear growth detectable by means of $\Delta$ for $\mu \leq 130$ which agrees with our previously determined value $\mu^* = 146 \pm 6$.

\subsection{Time of major disconnection: $t_c$}

\begin{figure}[]
\begin{center}
	\includegraphics[width=\columnwidth]{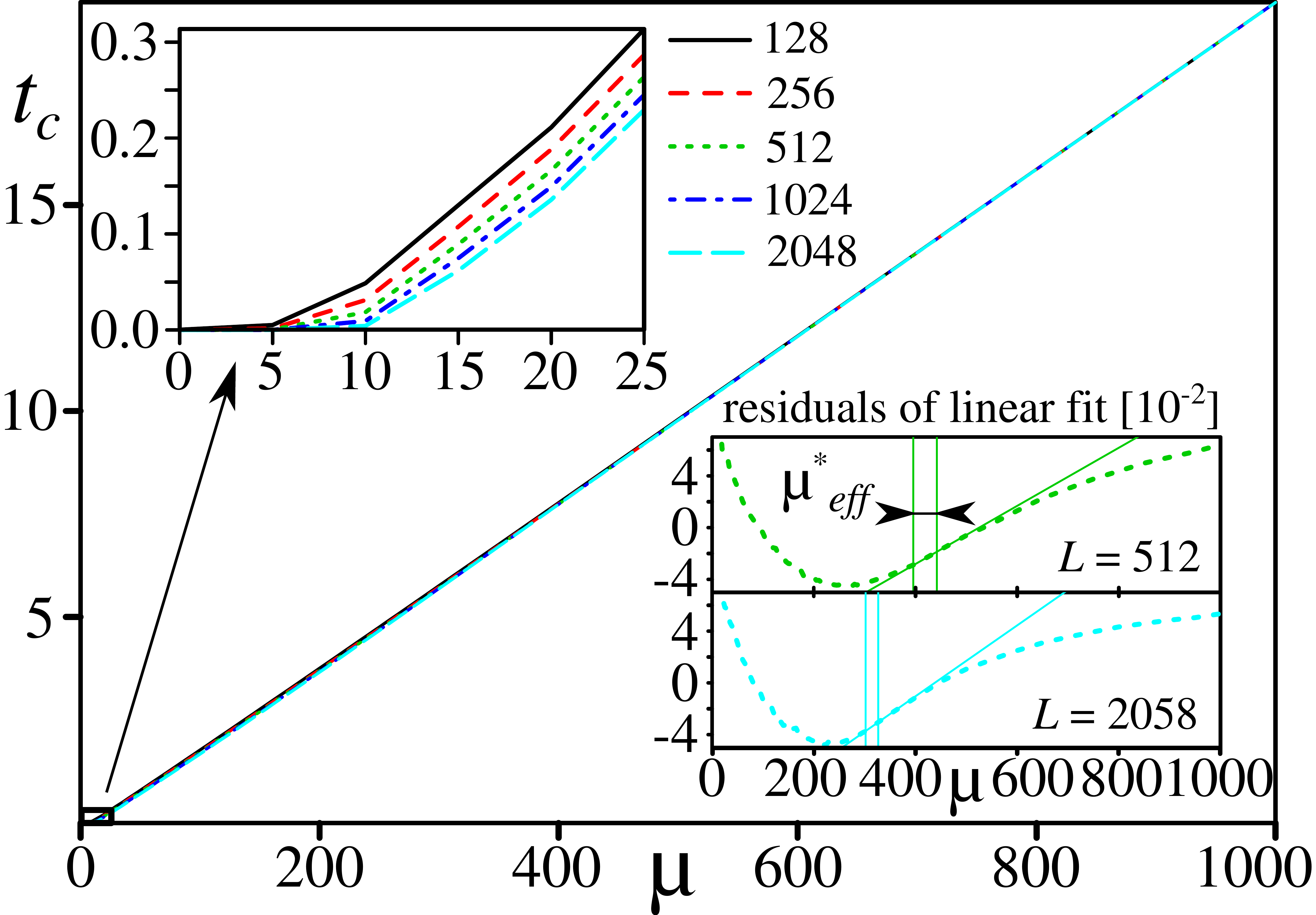}
\end{center}
\caption{
\label{fig:tc}
(Color online) \textbf{Time $t_c$ of the biggest jump $\Delta S$ due to a single avalanche.} $t_c$ is well fitted by a linear relation to $\mu$, except for very low $\mu$ (see the amplification in the inset at the upper left). Lower right inset: Not detectable by eye, the residuals of linear fits of $t_c$ of a particular system size ($L=\{512,2048\}$ as examples) reveal that the function $t_c(\mu)$ is super-linear below the transition failing threshold $\mu^*_{eff}$ and sub-linear above. Vertical lines bound the estimated range of $\mu^*_{eff}$ (compare Fig.\ \ref{fig:mustar}) which includes the inflection point (linear extrapolations from the estimation range help to verify that by eye). $t_c$ was measured at $\mu=\{0,5,\ldots,295,300,350,\ldots,950,1000\}$ and for $L=\{128,256,512,1024,2048\}$ the number of samples is $\{1600,800,400,200,100\}$. $\varepsilon=0.01$.
}
\end{figure}

The time $t_c$, at which the largest decrease in $S$ due to a single avalanche occurs (abrupt collapse for $\mu < \mu^*$), can be used to know when the network is connected ($t<t_c$) or disconnected ($t>t_c$). It is therefore of interest how $t_c$ scales with $\mu$ and $L$. We find that $t_c$ depends weakly on $L$ but is quite well fitted by a linear relation with $\mu$ (see Fig.\ \ref{fig:tc}). At first sight unnoticeable, the residuals of linear fits of $t_c$ for a particular system size reveal that for every $L$ the function $t_c(\mu)$ is super-linear below the effective critical threshold  $\mu^*_{eff}$ of that system size and sub-linear above. Since $\mu^*_{eff}$ varies with $L$ as seen in Fig.\ \ref{fig:mustar}, the dependence of $t_c$ on $L$ changes with $\mu$ and can not be described in a simple way. Note that we found for the range of system sizes considered here that for $\mu \gtrsim 200$ the difference between $t_c$ for different $L$ decreases with increasing $\mu$ in absolute value.

\subsection{Role of toppling threshold $z_c$ and mass increment $\Delta z$}

So far, we fixed the amount of mass $\Delta z=1$ added stepwise to the system and the toppling threshold $z_c=4$ of nodes. One might wonder what is the influence of these parameters. $\Delta z$, $z_c$, and the failing threshold $\mu$ are not independent, and so we decided to express everything in terms of $\Delta z$ and end up with the two adimensional variables $z_c/\Delta z$ and $\mu/\Delta z$, and measure time as $t/\Delta z$. Thus changing $\Delta z$ to $\Delta z' =k \Delta z$ has the same effect as leaving $\Delta z$ unchanged but choose $z_c'=z_c/k$ and $\mu'=\mu/k$.
For simplicity, we fix $\Delta z=1$ and only investigate the influence of $z_c$.

\begin{figure}[]
\begin{center}
	\includegraphics[width=\columnwidth]{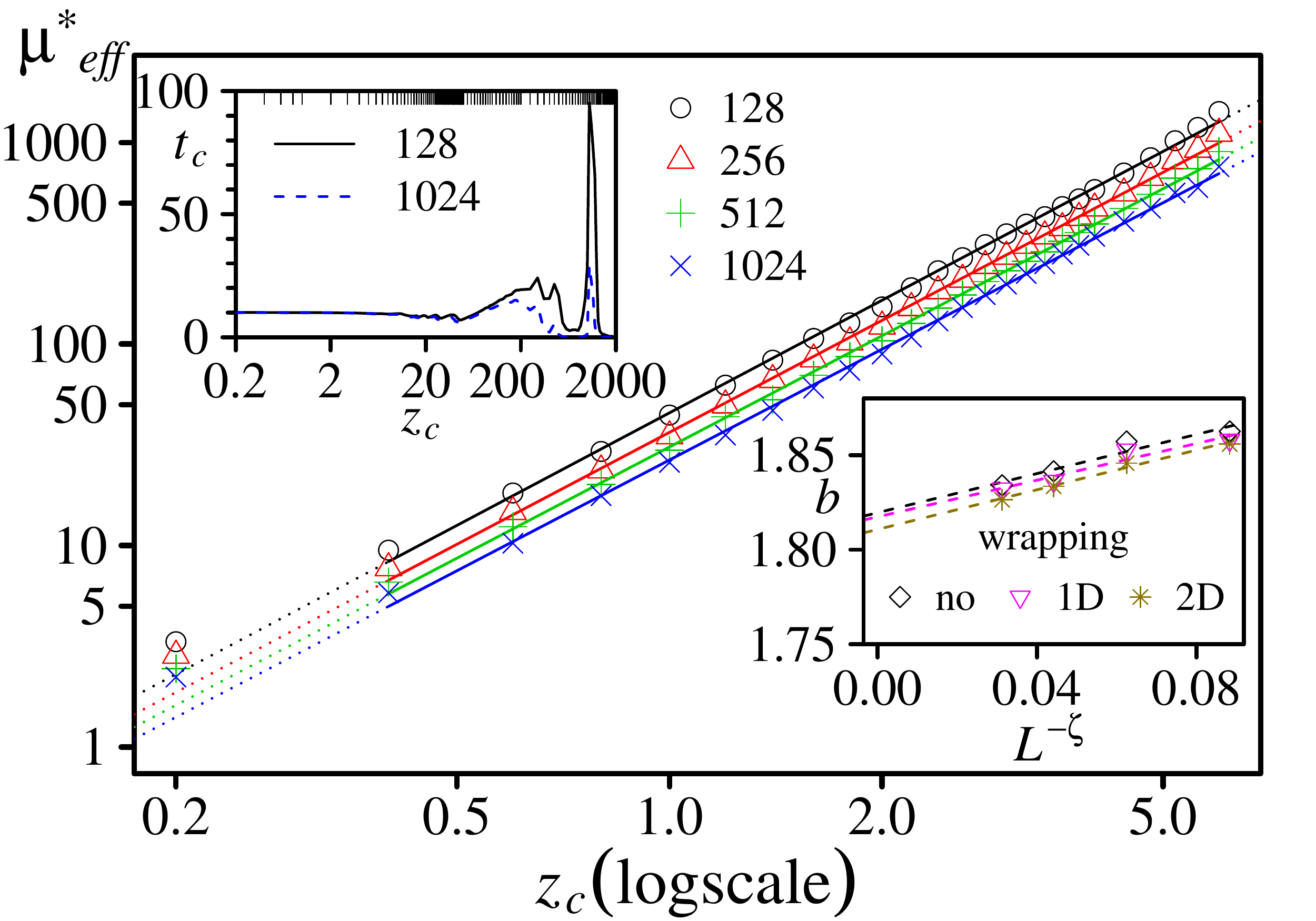}
\end{center}
\caption{
\label{fig:zc}
(Color online) \textbf{Influence of the toppling threshold $z_c$.} The critical failing threshold $\mu^*$ increases with $z_c$. Here we show one estimation of $\mu^*_{eff}$ (2D-wrapping, compare Fig.\ \ref{fig:mustar}) versus $z_c$. To find $\mu^*_{eff}$, 1000 samples each were measured in steps of $0.5 \le \Delta \mu \le 50$ depending on $\mu$. Error bars below $2\%$. Lower right inset: Finite size scaling of exponent $b$ of the power law fitting $\mu^*_{eff} \sim z_c^b$ for $0.4 \le z_c \le 6$. Error bars below the size of the symbols. With different estimations of $\mu^*_{eff}$ we predict $\mu^* \sim {z_c}^{b_c}$ with $b_c = 1.8 \pm 0.1$, under the assumption $(b-b_c) \sim L^{-\zeta}$ and $ 0.1 < \zeta < 1$. In the plot $\zeta = 0.5$. Upper left inset: $t_c$ versus $z_c$ for $\mu=500$ and $L=\{128,1024\}$ (number of samples is $\{100,50\}$, error bars below $1\%$). $t_c$ is nonsensitive to changes in $z_c$ for $z_c \lessapprox 5$, i.e., $z_c \ll \mu$. $\varepsilon = 0.01$.
}
\end{figure}

When analyzing the sensitivity of the effective thresholds $\mu^*_{eff}$, we find that $\mu^*_{eff} \sim {z_c}^b$ as seen in Fig.\ \ref{fig:zc} and with a size dependence analysis for $0.4 \le z_c \le 6$ (lower right inset of Fig.\ \ref{fig:zc}) we estimate that the critical threshold 
\begin{equation}\label{eq:mu_vs_zc}
\mu^* \sim {z_c}^{b_c}
\end{equation} with $b_c = 1.8 \pm 0.1$, obtained under the assumption $(b-b_c) \sim L^{-\zeta}$ and $ 0.1 < \zeta < 1$.

In general, we find that
\begin{equation}
\frac{\mu^*}{\Delta z} \sim \left(\frac{z_c}{\Delta z}\right)^{b_c},
\end{equation}
and thus,
\begin{equation}\label{eq:mu_vs_zc_and_Dz}
\mu^* \sim z_c^{b_c} {\Delta z}^{1-b_c}.
\end{equation}
To drive the system away from an abrupt collapse, one can, for example, lower $\mu^*$ while keeping $\mu$ constant. From Eq.\ (\ref{eq:mu_vs_zc_and_Dz}) we can directly see that since $b_c>1$, both decreasing $z_c$ and increasing $\Delta z$ will move the system away from an abrupt collapse.

The time $t_c$ is robust to changes in $z_c$ and $\Delta z$ for $z_c / \Delta z \lessapprox 5$, but not for larger $z_c / \Delta z$ as seen in the upper-left inset of Fig.\ \ref{fig:zc} for $\mu=500$ and $\Delta z =1$. For some very large $z_c$, where it only needs a few transportation through each link to fail it, the inhibiting effect of link failing can become relevant and therefore we can see an increase in $t_c$. However, this is a finite-size effect. To give an impression of the inhibiting effect of link failing, let us consider the value $z_c=1050$ where we find the highest peak (measurements in steps of $\Delta z_c=50$ around this point). In that case the mass transported through links is often such that it takes only two topplings from any of the two nodes connected together to fail the link. Thus many links often transport mass only once forth and back (often during the same avalanche) before they fail and therefore mass often ends up on isolated nodes. This can only prevent a devastating avalanche in small enough system sizes since with increasing system size the probability to overcome this inhibiting barrier at one point to start a devastating avalanche increases and therefore loses relevance on the large scale.

\subsection{Preventing an abrupt collapse}

To avoid an abrupt collapse it is crucial to reduce the empowering effect that damaging events have on avalanches. Increasing the failing threshold of links leads to larger fluctuations in usage at the time of failing and reduces the probability of simultaneous failing of neighboring links. We have seen how the toppling threshold and the mass increment added stepwise to the system influence its behavior and that both decreasing the toppling threshold and increasing the mass increment drive the system away from an abrupt collapse.

\begin{figure}[]
\begin{center}
	\includegraphics[width=\columnwidth]{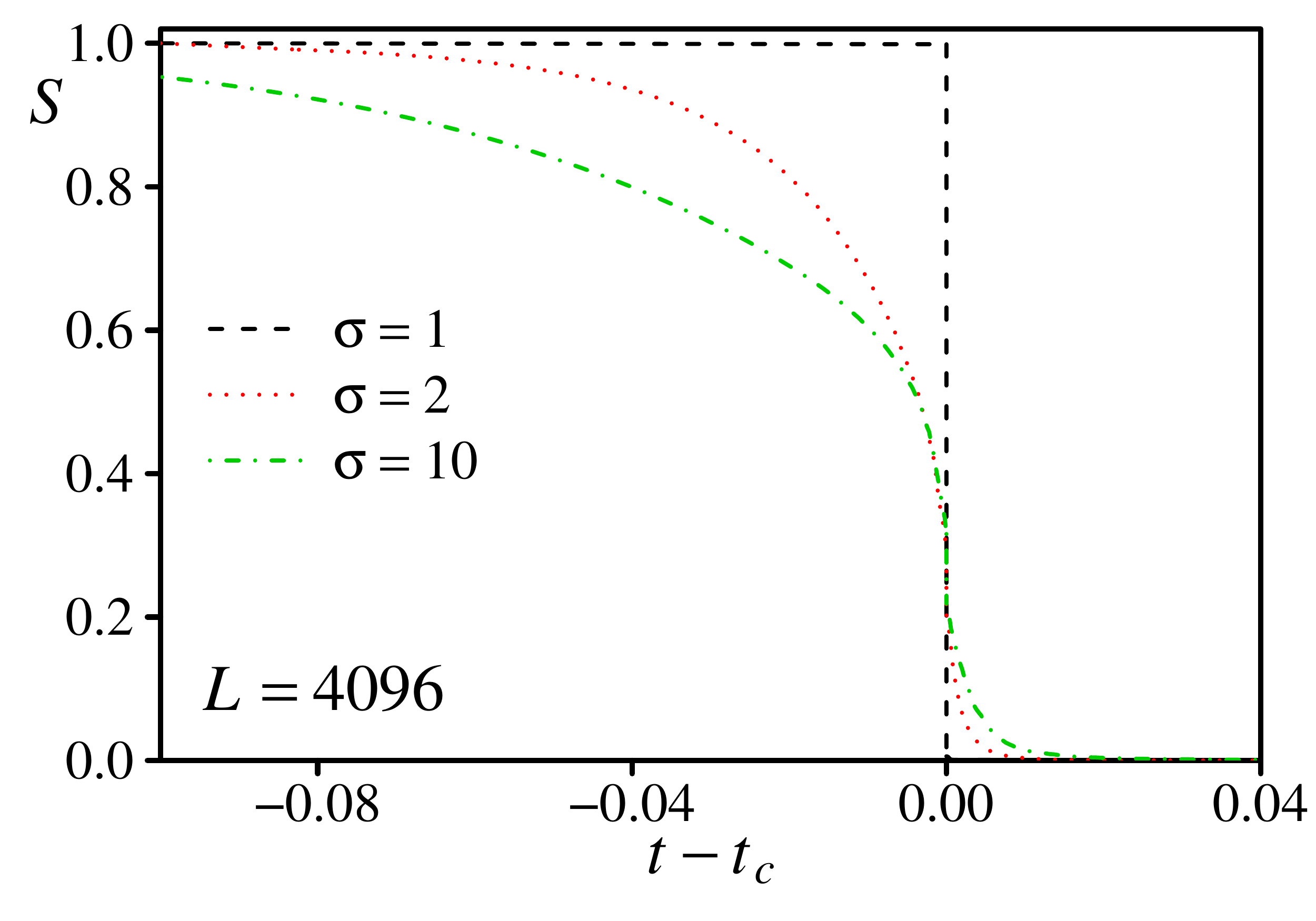}
\end{center}
\caption{
\label{fig:dispersion}
(Color online) \textbf{Dispersed failing thresholds can prevent an abrupt collapse.} Behavior of $S$ versus time for Gaussian distributed failing thresholds with mean $\mu=100$ and standard deviations $\sigma=\{1,2,10\}$ for $L=4096$ (averaged over 25 samples). An abrupt collapse is still seen for $\sigma=1$ but not for $\sigma=\{2,10\}$. $\varepsilon=0.01$.
}
\end{figure}

Another possibility to prevent an abrupt collapse is the use of a heterogeneous distribution of link thresholds. In Fig.\ \ref{fig:dispersion} we show that if one uses normally distributed failing thresholds with a mean of $\mu=100 < \mu^*$, we still observe an abrupt collapse for a small standard deviation of the links of $\sigma=1$ but already a gradual destruction for $\sigma=2$.
One can think of many other modifications which might also suppress a devastating avalanche such as limiting the transport capacity of links or the toppling outflow of nodes, introducing a progressive rate of dissipation, or letting links fail immediately during a transportation such that part of the outflow of the toppling node is sent back or is redistributed.

\section{Conclusions}\label{sec:Conclusions}

Although the size of cascades is limited by an exponential cutoff distribution in the absence of link failure, we observed that due to the positive feedback of link failures a macroscopic devastating avalanche occurs for failing thresholds below a critical level. 
To avoid a global catastrophe, it is therefore crucial to suppress positive feedback of failures. We showed with our model that not only dispersed failing thresholds can prevent an abrupt collapse but also to decrease the toppling threshold of the nodes and to increase the mass increment added stepwise to the system.

Our prediction of a lower bound for the critical threshold shows that when studying cascading phenomena in general, investigating not only the size and total damage of destructive events but also their step-wise evolution can be of valuable insight. That will also shed light on the behavior of supercritical events in large systems from simulations of rather small ones.

Future work might explore this model on other network topologies as for example networks obtained from real data.
As an extension, one could allow network elements to recover from usage or even rebuild them after failing to evaluate possible recovering policies.
Additionally, since often network elements fail due to extensive use during a short time, such as in electrical grids or the Internet, it would be important to investigate a version of the model where the failing threshold limits the allowed use of links or nodes per time interval, avalanche, or certain number of consecutive topplings.
If and how strong failures are self-sustaining will tell how different limitations trigger or suppress an abrupt collapse of the network and might lead to new strategies to suppress catastrophic events in networks in general.

\begin{acknowledgments}
We acknowledge financial support from the ETH Risk Center, the Brazilian institute INCT-SC, Grant No.\ FP7-319968-FlowCCS of the European Research Council (ERC) Advanced Grant, and the Portuguese Foundation for Science and Technology (FCT) under Contracts
No.\ EXCL/FIS-NAN/0083/2012, No.\ PEst-OE/FIS/UI0618/2014, and No.\ IF/00255/2013.
\end{acknowledgments}

\bibliography{bibliography_paper}

\end{document}